\documentclass[aip,jcp,reprint,onecolumn,amsmath,amssymb,amsfonts,showkeys]{revtex4-1}

\usepackage{graphicx,bm,microtype,hyperref,algpseudocode,subfigure,algorithm,algorithmicx,multirow,footnote,xcolor,physics,lipsum,amscd}

\usepackage[version=4]{mhchem}
\newcommand{\mc}{\multicolumn}

\newcommand{\bA}{\mathbf{A}}
\newcommand{\bB}{\mathbf{B}}

\newcommand{\bZ}{\mathbf{Z}}

\newcommand{\bO}{\mathbf{0}}

\newcommand{\ba}{\bm{a}}
\newcommand{\bb}{\bm{b}}
\newcommand{\br}{\bm{r}}

\DeclareMathOperator{\erfc}{erfc}

\newcommand{\alert}[1]{\textcolor{black}{#1}}

\newcommand{\sexpval}[1]{[ #1 ]}
\newcommand{\sbraket}[2]{[ #1 | #2 ]}
\newcommand{\smel}[3]{[ #1 | #2 | #3 ]}

\usepackage{tikz}
\usetikzlibrary{arrows,positioning,shapes.geometric}
\usetikzlibrary{decorations.pathmorphing}

\tikzstyle{level 1}=[level distance=5cm, sibling distance=4cm]
\tikzstyle{level 2}=[level distance=5cm, sibling distance=2cm]
\tikzstyle{level 3}=[level distance=5cm, sibling distance=2cm]
\tikzstyle{end} = [circle, minimum width=4pt,fill, inner sep=0pt]

\usepackage{tikz}
\usetikzlibrary{arrows,positioning,shapes.geometric}
\usetikzlibrary{decorations.pathmorphing}

\tikzset{snake it/.style={
decoration={snake, 
    amplitude = .4mm,
    segment length = 2mm},decorate}
}
    
\begin{document}

\title{Recurrence relations for four-electron integrals over Gaussian basis functions}
\author{Giuseppe M. J. Barca}
\affiliation{Research School of Chemistry, Australian National University, ACT 2601, Australia}
\author{Pierre-Fran\c{c}ois Loos}
\thanks{Corresponding author}
\email{loos@irsamc.ups-tlse.fr}
\affiliation{Laboratoire de Chimie et Physique Quantiques, Universit\'e de Toulouse, CNRS, UPS, France}
\affiliation{Research School of Chemistry, Australian National University, ACT 2601, Australia}

\begin{abstract}
In the spirit of the Head-Gordon-Pople algorithm, we report vertical, transfer and horizontal recurrence relations for the efficient and accurate computation of four-electron integrals over Gaussian basis functions.
Our recursive approach is a generalization of our algorithm for three-electron integrals [J.~Chem.~Theory Comput.~12, 1735 (2016)].
The RRs derived in the present study can be applied to a general class of multiplicative four-electron operators.
In particular, we consider various types of four-electron integrals that may arise in explicitly-correlated F12 methods.
\end{abstract}

\maketitle

\section{
\label{sec:intro}
Introduction}
In 1985, starting from the Hylleraas functional \cite{Hylleraas28, Hylleraas29, Sinagoglu62} and using the interelectronic distance $r_{12} = \abs{\br_1 - \br_2}$ as a correlation factor, Kutzelnigg derived the seminal form of the MP2-R12 equations. \cite{Kutzelnigg85} 
This explicitly-correlated method (more formally stated together with Klopper in 1987 \cite{Klopper87}) was later extended to higher level of theory and more accurate correlation factors $f_{12} = f(r_{12})$, \cite{Noga94, Klopper91a, Klopper91b, Termath91, Persson96, Persson97, May04, Tenno04a, Tew05, May05} such as Gaussian geminals 
\begin{equation}
	f_{12} = \exp( - \lambda r_{12}^2 ),
\end{equation}
or Slater geminals 
\begin{equation}
	f_{12} = \exp( - \lambda r_{12} ).
\end{equation}
The resulting ``F12 methods'' achieve chemical accuracy for small organic molecules with relatively small Gaussian basis sets \cite{Klopper06, Hattig12, Kong12, Tenno12a, Tenno12b, Gruneis17} and are quickly becoming the first-choice method for high accuracy. \cite{Tenno12a, Tenno12b} 

However, the inclusion of the correlation factor $f_{12}$ dredged up an old problem: in addition to two-electron integrals (traditional ones and new ones), three-electron integrals over $f_{13}f_{23}$, $r_{12}^{-1}f_{13}$ and $r_{12}^{-1}f_{13}f_{23}$, as well as four-electron integrals over $r_{12}^{-1}f_{14}f_{23}$, $r_{12}^{-1}f_{13}f_{34}$ and $r_{12}^{-1}f_{13}f_{14}$ arise.
Except when one uses Gaussian geminals, \cite{Boys60, Singer60} these integrals are not known analytically and, at that time, the only way to evaluate them would have been via expensive Gauss-Legendre quadratures. \cite{Preiskorn85, Clementi89}
Additionally, citing Kutzelnigg and Klopper, \cite{Kutzelnigg91} 
``\textit{even if fast procedures for the evaluation of these integrals were available, one would have to face the problem of the large number of these integrals; while that of two-electron integrals is $\sim N^{4}$, there are $\sim N^{6}$ three-electron and $\sim N^{8}$ four-electron integrals. 
The storing and manipulating of these integrals could be handled only for extremely small basis sets.}"

Undoubtedly, in the late 80's, the two-electron integrals technology was still in development. \cite{MD78, PH78, King76, Dupuis76, Rys83, OS1, HGP, Gill94b}
Nowadays, though still challenging, these integrals could be computed much more effectively via judicious recursive schemes, designing the quadrature only to the fundamental integrals. \cite{3ERI1}
Another important remark is that the actual number of \textit{significant} (i.e.~greater than a given threshold) three- and four-electron integrals in a large system, is, at worst, $\order{N^{3}}$ or $\order{N^{4}}$. 
These kinds of scaling are achievable, for example, by exploiting robust density fitting \cite{Womack2014} or upper bound-based screening methods. \cite{3ERI2} 

Nevertheless, the success of the R12 method was due to the decision of avoiding three- and four-electron integrals entirely through the insertion of the \textit{resolution of the identity} (RI) \cite{Kutzelnigg91, Hattig12, Werner07}
\begin{equation}
\label{eq:ri}
	\hat{I} \approx \sum_{\mu}^{N_\text{RI}}{\ket{\chi_{\mu}}\bra{\chi_{\mu}}}.
\end{equation}
In this way, three- and four-electron integrals are approximated as linear combinations of products of more conventional two-electron integrals.
Of course, the accuracy of the RI approximation \eqref{eq:ri} relies entirely on the assumption that the auxiliary basis set is sufficiently large ($N_\text{RI} \gg N$). 
Thus, in the general context of explicitly correlated methods, it is licit to ask: what is the suitable method \alert{to} evaluate three- and four-electron integrals?

It not clear to us that such method is RI. 
In fact, eschewing the RI approximation would offer at least two advantages: 
i) smaller one-electron basis as the larger auxiliary basis set would not be required anymore; 
ii) the three- and four-electron integrals would be computed exactly.
\alert{Moreover, one could avoid the commutator rearrangements involved in the computation of integrals over the kinetic energy operator. \cite{Rohse93}}

In a recent paper, \cite{3ERI1} we reported recurrence relations (RRs) to compute three-electron integrals over Gaussian basis functions for multiplicative general operators of the form $f_{12}g_{13}h_{23}$.
Here, we generalize our previous study to four-electron integrals.

The present paper is organized as follows.
In Sec.~\ref{sec:integrals}, we introduce notations and we define various key quantities.
Section~\ref{sec:FI} explains how to calculate fundamental integrals required to start the recursive scheme.
In Sec.~\ref{sec:RR}, we reports vertical, transfer and horizontal RRs for four-electron integrals.
Finally, in Sec.~\ref{sec:algo}, we propose a recursive scheme based on these RRs to calculate classes of four-electron integrals.
Atomic units are used throughout. 

\section{
\label{sec:integrals}
Four-electron integrals}
A primitive Gaussian-type function (PGF) is specified by an orbital exponent $\alpha$, a center $\bA=(A_x,A_y,A_z)$, and angular momentum $\ba=(a_{x},a_{y},a_{z})$:
\begin{equation}
\label{eq:def1}
	\varphi_{\ba}^{\bA}(\bm{r})  	
	= (x-A_{x})^{a_{x}} (y-A_{y})^{a_{y}} (z-A_{z})^{a_{z}} e^{-\alpha \left| \br-\bA \right|^2}.
\end{equation}
A contracted Gaussian-type function (CGF) is defined as a normalized sum of PGFs
\begin{equation}
\label{eq:def1b}
	\psi_{\ba}^{\bA}(\br)	
	=\sum_{k=1}^{K_A} D_{\ba k} (x-A_{x})^{a_{x}} (y-A_{y})^{a_{y}} (z-A_{z})^{a_{z}} e^{-\alpha_k \left| \br-\bA \right|^2},
\end{equation}
where $K_A$ is the degree of contraction and the $D_{\ba k}$ are contraction coefficients.
Throughout this paper, we use physicists notations, and we write the integral over a four-electron operator $f_{1234}$ of CGFs as
\begin{widetext}
\begin{equation}
\begin{split}
\label{eq:def2}
	\braket{\ba_1 \ba_2 \ba_3 \ba_4}{\bb_1 \bb_2 \bb_3 \bb_4}
	& \equiv \mel{\ba_1 \ba_2 \ba_3 \ba_4}{f_{1234}}{\bb_1 \bb_2 \bb_3 \bb_4}
	\\
	& = \iiint 
	\psi_{\ba_1}^{\bA_1}(  \br_{1}) \psi_{\ba_2}^{\bA_2}(\br_{2}) \psi_{\ba_3}^{\bA_3}(\br_{3})\psi_{\ba_4}^{\bA_4}(\br_{4})
	f_{1234}
	\psi_{\bb_1}^{\bB_1}(  \br_{1}) \psi_{\bb_2}^{\bB_2}(\br_{2}) \psi_{\bb_3}^{\bB_3}(\br_{3})\psi_{\bb_4}^{\bB_4}(\br_{4})
	d \br_{1} d \br_{2} d \br_{3} d \br_{4}.
\end{split}
\end{equation}
Additionally, square-bracketed integrals denote integrals over PGFs:
\begin{equation}
\begin{split}
\label{eq:def2b}
	\sbraket{\ba_1 \ba_2 \ba_3 \ba_4}{\bb_1 \bb_2 \bb_3 \bb_4}
	& \equiv \smel{\ba_1 \ba_2 \ba_3 \ba_4}{f_{1234}}{\bb_1 \bb_2 \bb_3 \bb_4}
	\\
	& = \iiint 
	\varphi_{\ba_1}^{\bA_1}(  \br_{1}) \varphi_{\ba_2}^{\bA_2}(\br_{2}) \varphi_{\ba_3}^{\bA_3}(\br_{3}) \varphi_{\ba_4}^{\bA_4}(\br_{4})
	f_{1234}
	\varphi_{\bb_1}^{\bB_1}(  \br_{1}) \varphi_{\bb_2}^{\bB_2}(\br_{2}) \varphi_{\bb_3}^{\bB_3}(\br_{3}) \varphi_{\bb_4}^{\bB_4}(\br_{4})
	d \br_{1} d \br_{2} d \br_{3} d \br_{4}.
\end{split}
\end{equation}
The fundamental integral (i.e.~the integral in which all eight basis functions are $s$-type PGFs) is defined as $\sexpval{\bO} \equiv \sbraket{\bO \bO \bO \bO}{\bO \bO \bO \bO} $ with $\bO=(0,0,0)$.
The Gaussian product rule reduces it from eight to four centers:
\begin{equation}
\label{eq:def4}
	\sexpval{\bO} = 
	S_{1} S_{2} S_{3} S_{4}
	\iiint \varphi_{\bO}^{\bZ_1}(\bm{r}_{1}) \varphi_{\bO}^{\bZ_2}(\bm{r}_{2}) \varphi_{\bO}^{\bZ_3}(\bm{r}_{3}) \varphi_{\bO}^{\bZ_4}(\bm{r}_{4})
	f_{1234} d \bm{r}_{1} d \bm{r}_{2} d \bm{r}_{3} d \bm{r}_{4},
\end{equation}
\end{widetext}
where 
\begin{align}
	\zeta_i & = \alpha_i + \beta_i,		
	& 
	\bZ_i & = \frac{\alpha_i \bA_i + \beta_i \bB_i}{\zeta_i},		
	& 
	S_{i} & = \exp(-\frac{\alpha_i \beta_i}{\zeta_i} \abs{\bA_i\bB_i}^2),
\end{align}
with \alert{$\bA_i\bB_i = \bA_i - \bB_i$}.
For conciseness, we will adopt a notation in which missing indices represent $s$-type Gaussians.  
For example, $\sexpval{\ba_2\ba_3}$ is a shorthand for $\sexpval{\bO\ba_2\ba_3\bO | \bO\bO\bO\bO}$.  
We will also use unbold indices, e.g. $\sexpval{a_1a_2a_3a_4|b_1b_2b_3b_4}$ to indicate a complete class of integrals from a shell-octet.

\begin{figure*}
	\includegraphics[width=\linewidth]{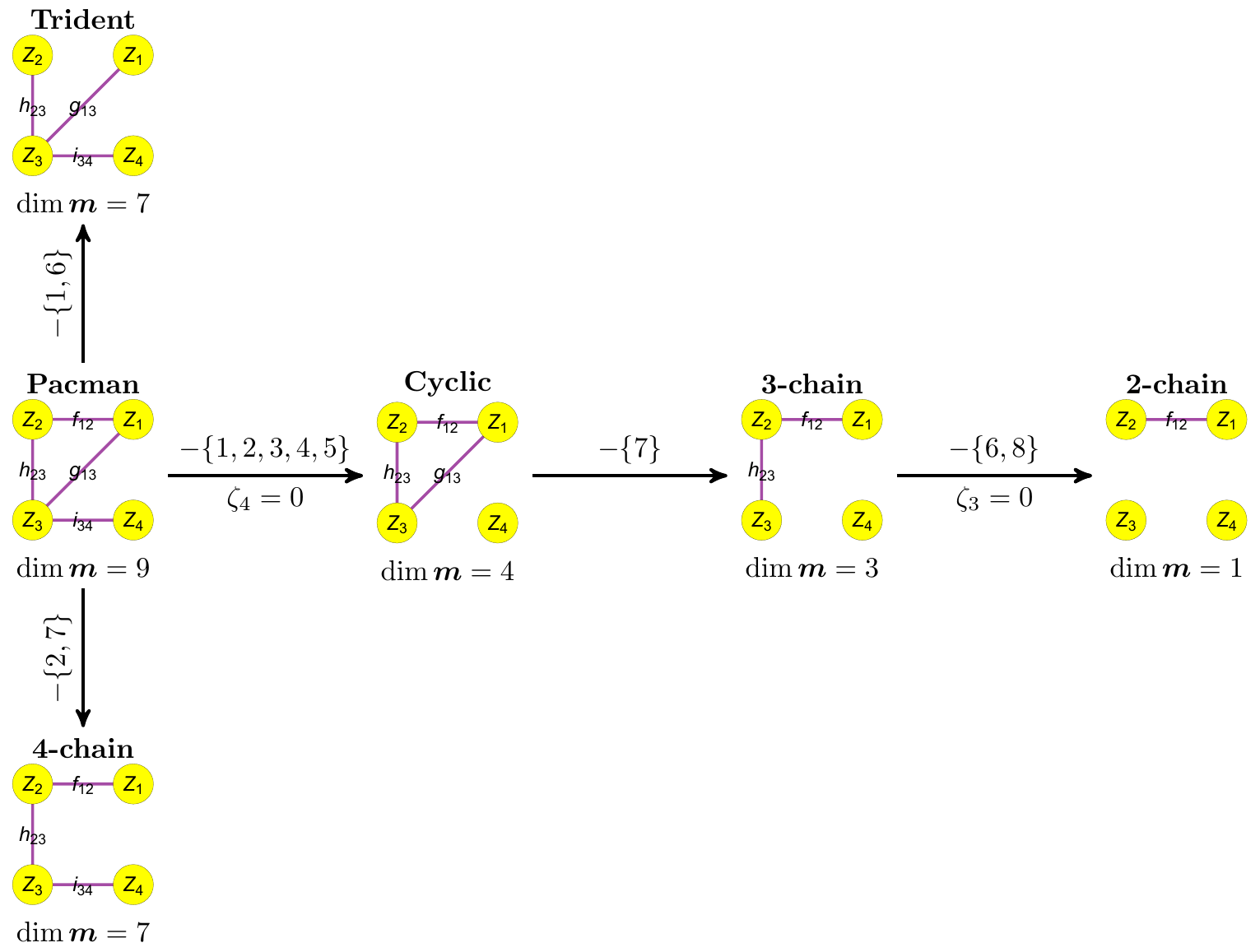}
\caption{
\label{fig:tree}
Diagrammatic representation of various four-, three- and two-electron integrals involved in explicitly-correlated methods.
\alert{$\dim \bm{m} $ refers to the dimensionality of the auxiliary index $\bm{m}$ (see Eq.~\eqref{eq:fi6}).
The values in the curly brackets indicates which components of the auxiliary index vector $\bm{m}$ must be removed.}}
\end{figure*}

\subsection{
\label{sec:operators}
Four-electron operators}
In the present study, we are particularly interested in the four-electron operators $g_{13}h_{23}i_{34}$ (trident) and $f_{12}h_{23}i_{34}$ (four-electron chain or 4-chain) because they can be required in explicitly-correlated methods such as F12 methods. \cite{Tenno12a, Tenno12b, Kong12, Hattig12}
Explicitly-correlated calculations may also \alert{require} three-electron integrals over the (3-chain) $f_{12}h_{23}$ and (cyclic) $f_{12}g_{13}h_{23}$ operators, as well as two-electron integrals over $f_{12}$.
However, we will eschew the study of the two-electron integrals here as they have been extensively studied in the past 25 years. \cite{Kutzelnigg91, Klopper92, Persson97, Klopper02, Manby03, Werner03, Klopper04, Tenno04a, Tenno04b, May05, Manby06, Tenno07, Komornicki11, Reine12}
Note that the nuclear attraction integrals can be easily obtained by taking the large-exponent limit of a $s$-type shell-pair.
\alert{We refer the interested reader to Refs.~\onlinecite{PRISM91, Gill94b} for more details about the computation of nuclear attraction integrals.}

The structure of these operators is illustrated in Fig.~\ref{fig:tree}, where we have adopted a diagrammatic representation.
Starting with the ``pacman'' operator $f_{12} g_{13}h_{23}i_{34}$, we are going to show that one can easily derive all the RRs required to compute two-, three- and four-electron integrals following simple rules.
Therefore, in the following, we will focus our analysis on this master ``pacman'' operator.

\begin{table}
\caption{
\label{tab:kern}
Kernels $F(t)$ of the Gaussian integral representation for various $f_{12}$ operators. 
$\delta(x)$ and $\theta(x)$ are respectively the Dirac delta and Heaviside step functions, and $\erf(x)$ and $\erfc(x)$ are the error function and its complement version, respectively. \cite{NISTbook}} 
\begin{tabular}{lc}
\hline
\hline
$f_{12}$								&	$F(t)$ 									\\
\hline
 $1$									&	$\delta(t)$									\\[5pt]
 $r_{12}^{-1}$							&	$2/\sqrt{\pi}$								\\[5pt]
 $r_{12}^{-2}$							&	$2t$										\\[5pt]
 $(r_{12}^2+\lambda^2)^{-1/2}$				&	$(2/\sqrt{\pi}) \exp(-\lambda^2 t^2)$				\\[5pt]
 $\exp (-\lambda\,r_{12} )$					&	$(\lambda\,t^{-2}/\pi) \exp(-\lambda^2 t^{-2}/4)$		\\[5pt]
 $r_{12}^{-1} \exp (-\lambda\,r_{12})$			&	$(2/\sqrt{\pi}) \exp(-\lambda^2 t^{-2}/4)$			\\[5pt]
 $\exp(-\lambda^2 r_{12}^2 )$				&	$\delta(t-\lambda)$							\\[5pt]
 $r_{12}^{-1} \erfc (\lambda\,r_{12})$			&	$(2/\sqrt{\pi})\,\theta(t-\lambda) $				\\[5pt]
 $r_{12}^{-1} \erf (\lambda\,r_{12})$			&	$(2/\sqrt{\pi}) \left[1-\theta(t-\lambda)\right] $		\\[5pt]
\hline
\hline
\end{tabular}	
\end{table}

\section{
\label{sec:FI}
Fundamental integrals}
The first step required to compute integrals of arbitrary angular momentum is the computation of the (momentumless) fundamental integrals $\sexpval{ \bO }$.
These are derived starting from Eq.~\eqref{eq:def4} using the Gaussian integral representation of each two-electron operator.
For instance, we have
\begin{equation}
\label{eq:fi2}
	f_{12} = \int_0^\infty F(t_{12})  \exp(-t_{12}^2 r_{12}^2) \,dt_{12},
\end{equation}
where $F(t_{12})$ is a Gaussian kernel. 
Table \ref{tab:kern} contains kernels $F(t)$ for a variety of important two-electron operators $f_{12}$.
\alert{From the formulas in Table \ref{tab:kern}, one can also easily deduce the kernels for related functions, such as $f_{12}^2$, $f_{12}/r_{12}$, and $\nabla^2 f_{12}$, which are of interest in explicitly-correlated methods. More general kernels can be found in Ref.~\onlinecite{3ERI1}.}

Next, the integration over $\bm{r}_{1}$, $\bm{r}_{2}$, $\bm{r}_{3}$ and $\bm{r}_{4}$ can be carried out, yielding 
\begin{equation}
\label{eq:fi3}
		\sexpval{\bO} =  S_{1} S_{2} S_{3} S_{4} 
		 \iiiint F(t_{12})\,  G(t_{13})\, H(t_{23})\, I(t_{34})\, w_{\bO}(\bm{t})  \,d\bm{t},
\end{equation}
where $\bm{t} = (t_{12},t_{13},t_{23},t_{34})$ and
\begin{equation}
\label{eq:fig0} 
	w_{\bO}(\bm{t}) =  \qty(\frac{\pi^4}{D(\bm{t})} )^{3/2}  \exp [ -\frac{N(\bm{t})}{D(\bm{t})} ].
\end{equation} 
Defining the following polynomials
\begin{subequations}
\begin{align}
	s_1 & = t_{12}^2,
	&
	s_2 & = t_{13}^2,
	&
	s_3 & = t_{23}^2,
	\\
	s_4 & = t_{34}^2,
	&
	s_5 & = t_{12}^2 t_{13}^2 + t_{12}^2 t_{23}^2 + t_{13}^2 t_{23}^2,
	&
	s_6 & = t_{12}^2 t_{34}^2,
	\\
	s_7 & = t_{13}^2 t_{34}^2,
	&
	s_8 & = t_{23}^2 t_{34}^2, 
	&
	s_9 & = t_{12}^2 t_{13}^2 t_{34}^2 + t_{12}^2 t_{23}^2 t_{34}^2 + t_{13}^2 t_{23}^2 t_{34}^2,
\end{align}
\end{subequations}
we have
\begin{subequations}
\begin{align}
	D(\bm{t})	 & = 
	\zeta _1 \zeta _2 \zeta _3 \zeta _4
	+ \left(\zeta _1+\zeta _2\right) \zeta _3 \zeta _4 s_1 
	+ \zeta _2 \zeta _4 \left(\zeta _1+\zeta _3\right) s_2
	+ \zeta _1 \zeta _4 \left(\zeta _2+\zeta _3\right) s_3
	+ \zeta_1 \zeta _2 \left(\zeta _3+\zeta _4\right) s_4	\notag
	\\
	& 
	+  \zeta _4 \left(\zeta _1+\zeta _2+\zeta _3\right) s_5
	+ \left(\zeta _1 +\zeta _2\right)\left( \zeta _3+\zeta _4\right) s_6
	+  \zeta_2 \left(\zeta _1+\zeta _3+\zeta _4\right) s_7
	+ \zeta _1 \left(\zeta _2+\zeta _3+\zeta _4\right) s_8 
	\\
	& 
	+ \left(\zeta _1+\zeta _2+\zeta _3+\zeta _4\right) s_9, \notag
	\label{eq:fi4} 	
	\\
	N(\bm{t}) 	& =
	\zeta _3 \zeta _4 \kappa _{12} s_1
	+ \zeta _2 \zeta _4 \kappa _{13} s_2
	+ \zeta _1 \zeta _4 \kappa _{23} s_3
	+ \zeta _1 \zeta _2 \kappa _{34} s_4
	+ \zeta _4 \left(\kappa _{12}+\kappa _{13}+\kappa _{23}\right) s_5 \notag
	\\
	& 
	+ \left[\left(\zeta _3+\zeta _4\right) \kappa_{12}+\left(\zeta _1+\zeta _2\right) \kappa _{34}\right] s_6 
	+ \zeta _2 \left(\kappa _{13}+\kappa _{14}+\kappa _{34}\right) s_7
	+ \zeta _1 \left(\kappa_{23}+\kappa _{24}+\kappa _{34}\right) s_8 
	\\
	& 
	+ \left(\kappa _{12}+\kappa _{13}+\kappa_{14}+\kappa _{23}+\kappa _{24}+\kappa _{34}\right) s_9,	 \notag
\end{align}
\end{subequations}
where \alert{$\kappa_{ij} = \zeta_i \zeta_j \,\abs{\bZ_i - \bZ_j}^2$}.

Following Obara and Saika, \cite{OS1, OS2} vertical RRs (VRRs) are obtained by differentiation of Eq.~\eqref{eq:fi3} with respect to the center coordinates.
Therefore, one can show that \eqref{eq:fi3} has to be generalized to the following form:
\begin{equation}
\label{eq:fi6}
		\sexpval{\bO}^{\bm{m}}
		= S_{1} S_{2} S_{3} S_{4}  
		\iiiint F(t_{12})\,  G(t_{13})\, H(t_{23})\, I(t_{34})\, w_{\bm{m}}(\bm{t})  \,d\bm{t},
\end{equation}
where $\bm{m}=(m_1,m_2,m_3,m_4,m_5,m_6,m_7,m_8,m_9)$ is, for the pacman operator, a nine-dimensional auxiliary index, and
\begin{equation}
	w_{\bm{m}}(\bm{t})=
	w_{\bO}(\bm{t}) \prod_{k=1}^{\dim{\bm{m}}} \qty[ \frac{s_k}{D(\bm{t})} ]^{m_k}.
\end{equation} 
As reported in Fig.~\ref{fig:tree}, while the fundamental integrals of the pacman operator contains 9 auxiliary indices (i.e.~$\dim \bm{m} = 9$), the two interesting four-electron operators (trident and chain) contains 7.
The cyclic and chain three-electron operators have only four and three, respectively, while the two-electron chain operator has a single $m$ component (i.e.~$\dim \bm{m} = 1$).
These numbers are drastically reduced if one uses Gaussian geminals due to their factorization properties. \cite{IntF12}

\section{
\label{sec:RR}
Recurrence relations}
In this Section, we report vertical, transfer and horizontal RRs for the computation of four-electron integrals of arbitrary angular momentum.
In particular, we refer the interested readers to the appendix of Ref.~\onlinecite{3ERI1} for more details about how to derive these VRRs \`a la Ahlrichs. \cite{Ahlrichs06}

\subsection{
\label{sec:VRR}
Vertical recurrence relations}
To build angular momentum over center $\bA_{1}$, we have derived the following 24-term VRR:
\begin{equation}
\begin{split}
\label{eq:VRR4A1}
	\sexpval{\ba_{1}^+ \ba_2 \ba_3 \ba_4}^{\bm{m}} 
	& = \bZ_1 \bA_1 \sexpval{\ba_1 \ba_2 \ba_3 \ba_4}^{\bm{m}} 
	- \zeta_2 \zeta_3 \zeta_4 \bZ_{12} \sexpval{\ba_1 \ba_2 \ba_3 \ba_4}^{\{1\}}
	- \zeta_2 \zeta_3 \zeta_4 \bZ_{13} \sexpval{\ba_1 \ba_2 \ba_3 \ba_4}^{\{2\}} 
	\\
	& - \zeta_4 ( \zeta_2 \bZ_{12} + \zeta_3 \bZ_{13} ) \sexpval{\ba_1 \ba_2 \ba_3 \ba_4}^{\{5\}}
	- \zeta_2 ( \zeta_3 + \zeta_4 ) \bZ_{12} \sexpval{\ba_1 \ba_2  \ba_3  \ba_4}^{\{6\}}
	\\
	& - \zeta_2 ( \zeta_3 \bZ_{13} + \zeta_4 \bZ_{14} ) \sexpval{\ba_1 \ba_2 \ba_3 \ba_4}^{\{7\}}
	- ( \zeta_2 \bZ_{12} + \zeta_3 \bZ_{13} + \zeta_4 \bZ_{14} ) \sexpval{\ba_1 \ba_2 \ba_3 \ba_4}^{\{9\}}
	\\
	& + \frac{\ba_1}{2\zeta_1} \Big\{
	\sexpval{\ba_1^- \ba_2 \ba_3 \ba_4}^{\bm{m}} 
	- \zeta_2 \zeta_3 \zeta_4 \bZ_{12} \sexpval{\ba_1^- \ba_2 \ba_3 \ba_4}^{\{1\}}
	- \zeta_2 \zeta_3 \zeta_4 \bZ_{13} \sexpval{\ba_1^- \ba_2 \ba_3 \ba_4}^{\{2\}} 
	\\
	& - \zeta_4 ( \zeta_2 + \zeta_3 ) \sexpval{\ba_1^- \ba_2 \ba_3  \ba_4}^{\{5\}}
	- \zeta_2 ( \zeta_3 + \zeta_4 ) \sexpval{\ba_1^- \ba_2 \ba_3  \ba_4}^{\{6\}}
	\\
	& - \zeta_2 ( \zeta_3 + \zeta_4 ) \sexpval{\ba_1^- \ba_2 \ba_3 \ba_4}^{\{7\}}
	- ( \zeta_2 + \zeta_3 + \zeta_4 ) \sexpval{\ba_1^- \ba_2 \ba_3 \ba_4}^{\{9\}}	
	\Big\} 
	\\
	& + \frac{\ba_2}{2} \Big\{
	\zeta_3 \zeta_4 \sexpval{\ba_1 \ba_2^- \ba_3 \ba_4}^{\{1\}}
	+ \zeta_4 \sexpval{\ba_1 \ba_2^-  \ba_3  \ba_4}^{\{5\}}
	+ ( \zeta_3 + \zeta_4 ) \sexpval{\ba_1 \ba_2^- \ba_3 \ba_4}^{\{6\}}
	+ \sexpval{\ba_1 \ba_2^- \ba_3 \ba_4}^{\{9\}}	
	\Big\} 
	\\
	& + \frac{\ba_3}{2} \Big\{
	\zeta_2 \zeta_4 \sexpval{\ba_1 \ba_2 \ba_3^- \ba_4}^{\{2\}}
	+ \zeta_4 \sexpval{\ba_1 \ba_2  \ba_3^-  \ba_4}^{\{5\}}
	+ \zeta_2 \sexpval{\ba_1 \ba_2 \ba_3^- \ba_4}^{\{7\}}
	+ \sexpval{\ba_1 \ba_2 \ba_3^- \ba_4}^{\{9\}}	
	\Big\} 
	\\
	& + \frac{\ba_4}{2} \Big\{
	\zeta_2 \sexpval{\ba_1 \ba_2 \ba_3 \ba_4^-}^{\{7\}}
	+ \sexpval{\ba_1 \ba_2 \ba_3 \ba_4^-}^{\{9\}}	
	\Big\},
	\end{split}
\end{equation}
where the superscript $+$ or $-$ denotes an increment or decrement of one unit of Cartesian angular momentum. (Thus, $\ba^{\pm}$ is analogous to $\ba \pm \bm{1}_i$ in the notation of Obara and Saika.) The value in the curly superscript indicates which component of the auxiliary index vector $\bm{m}$ is incremented.

Because Eq.~\eqref{eq:VRR4A1} builds angular momemtum over $\bA_{1}$ and all four bra centers have non-zero angular momentum, we will call this expression VRR$_{4}^{\bA_1}$.
The VRRs used to obtain $\sexpval{\ba_{1}^+ \ba_2 \ba_3}^{\bm{m}}$, $\sexpval{\ba_{1}^+ \ba_2}^{\bm{m}}$ and $\sexpval{\ba_{1}^+}^{\bm{m}}$ can be easily derived from Eq.~\eqref{eq:VRR4A1} by setting successively $\ba_4 = \bO$,  $\ba_3 = \bO$ and $\ba_2 = \bO$. These are respectively named VRR$_{3}^{\bA_1}$, VRR$_{2}^{\bA_1}$ , VRR$_{1}^{\bA_1}$.

One can easily derive VRR$_4$ for the trident and 4-chain operators following the simple rules given in Fig.~\ref{fig:tree}.
We obtain the VRRs for the trident operator by removing the terms $\{1\}$ and $\{6\}$. 
Similarly, 4-chain VRRs are obtained by removing the terms $\{2\}$ and $\{7\}$ .
This yields a 18- and 17-term VRR$_{4}^{\bA_1}$ for the trident and 4-chain operators, respectively.

\begin{figure*}
	\includegraphics[width=0.4\textwidth]{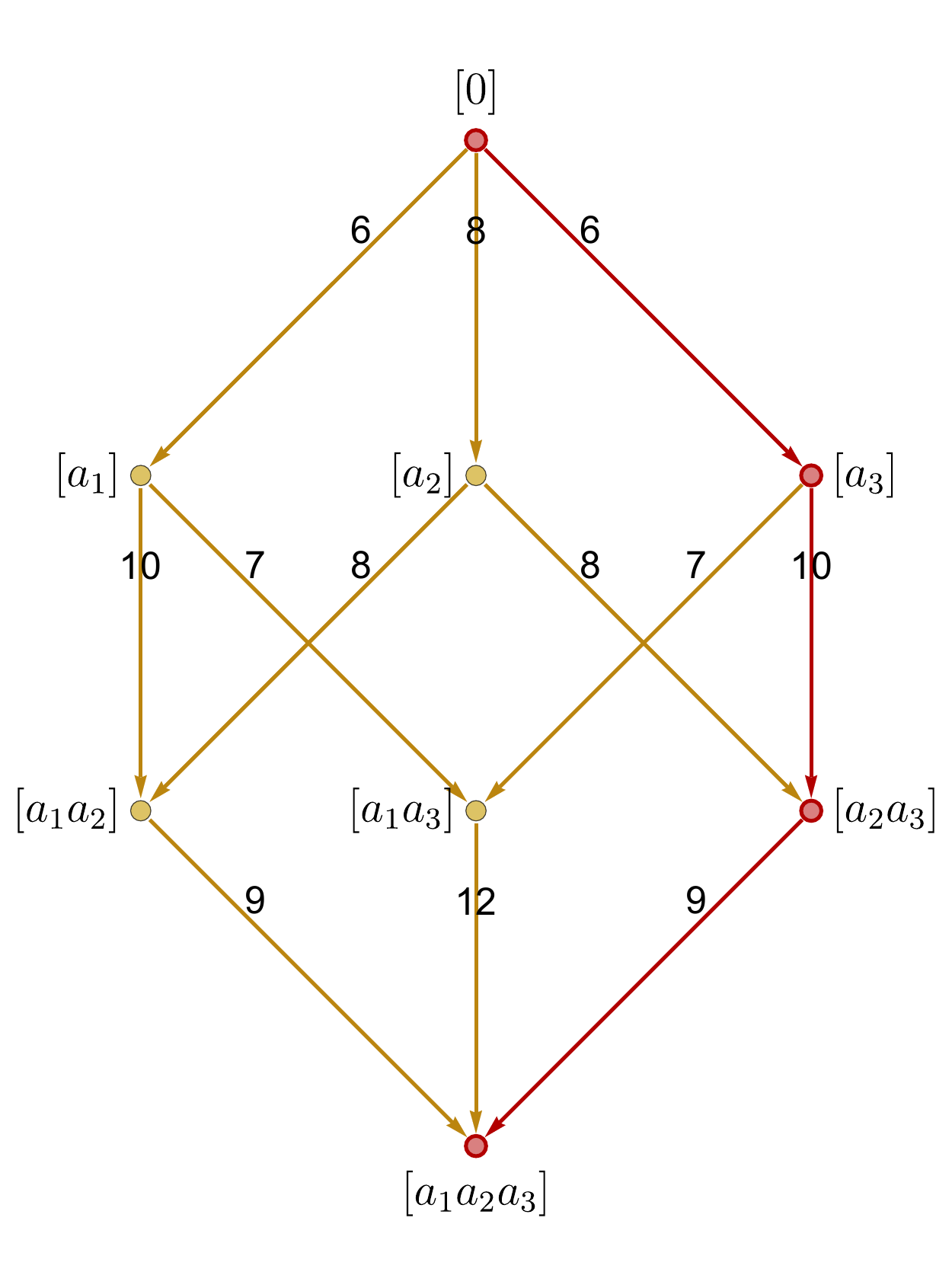}
	\includegraphics[width=0.4\textwidth]{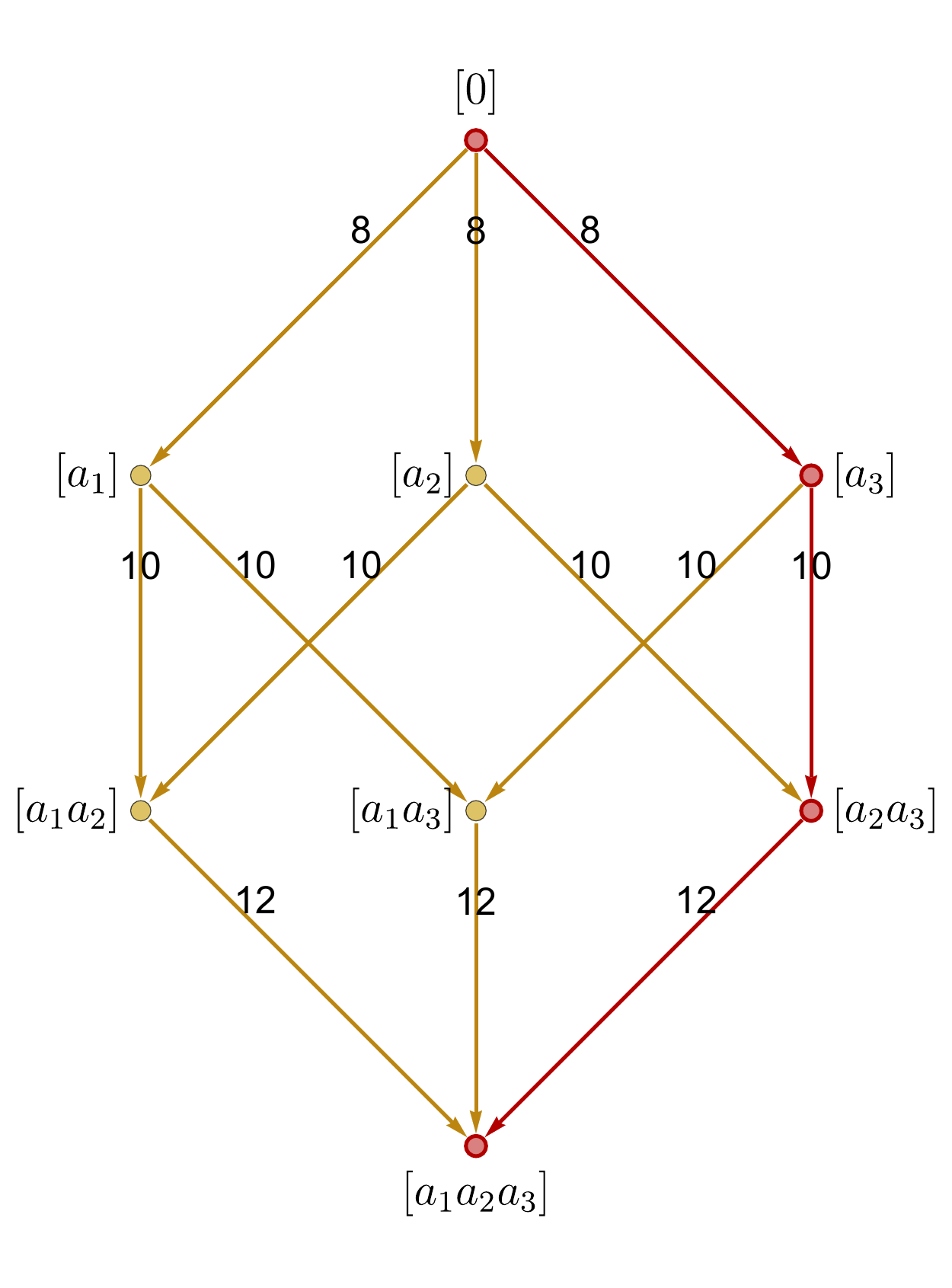}
	\includegraphics[width=0.49\textwidth]{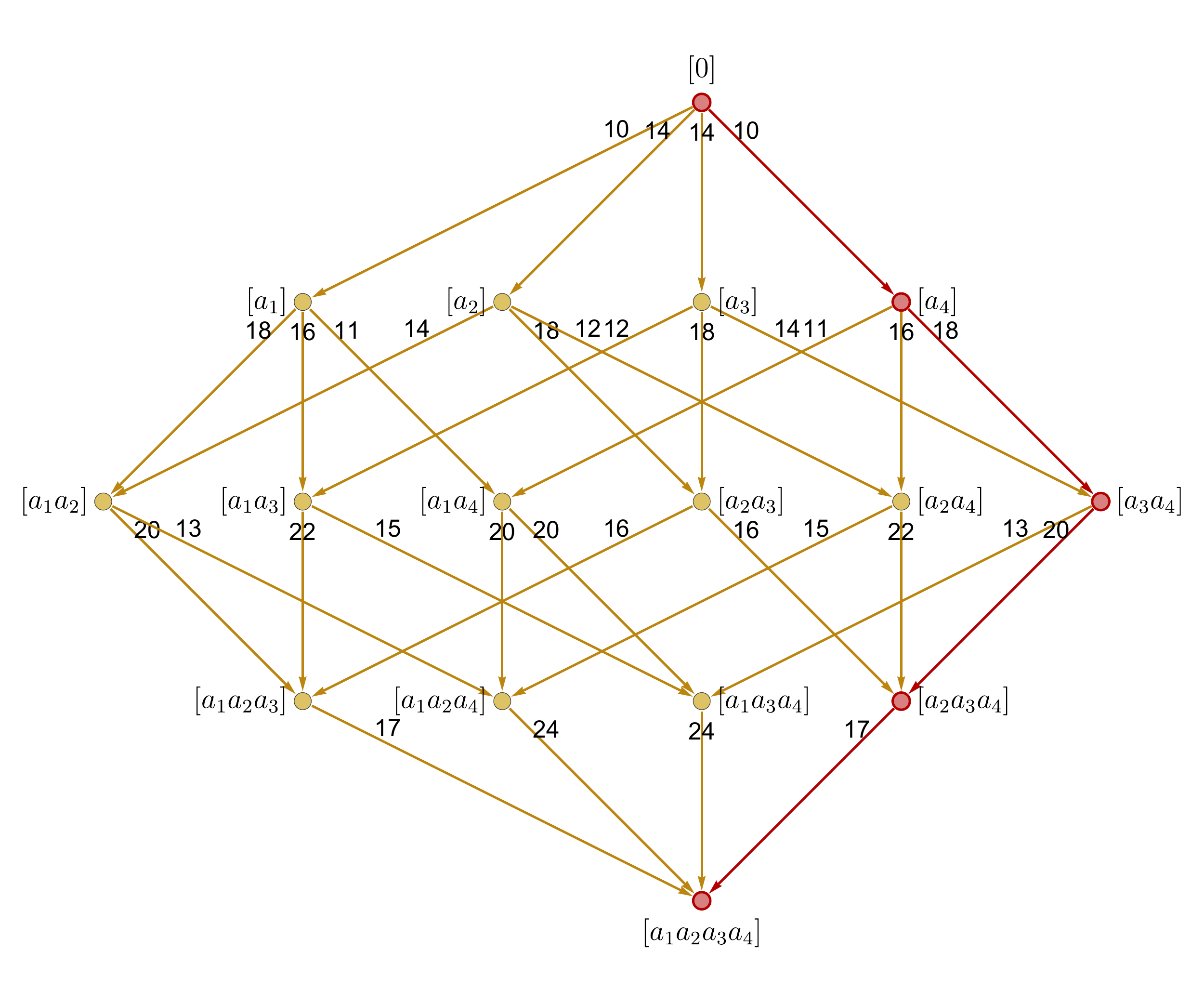}
	\includegraphics[width=0.49\textwidth]{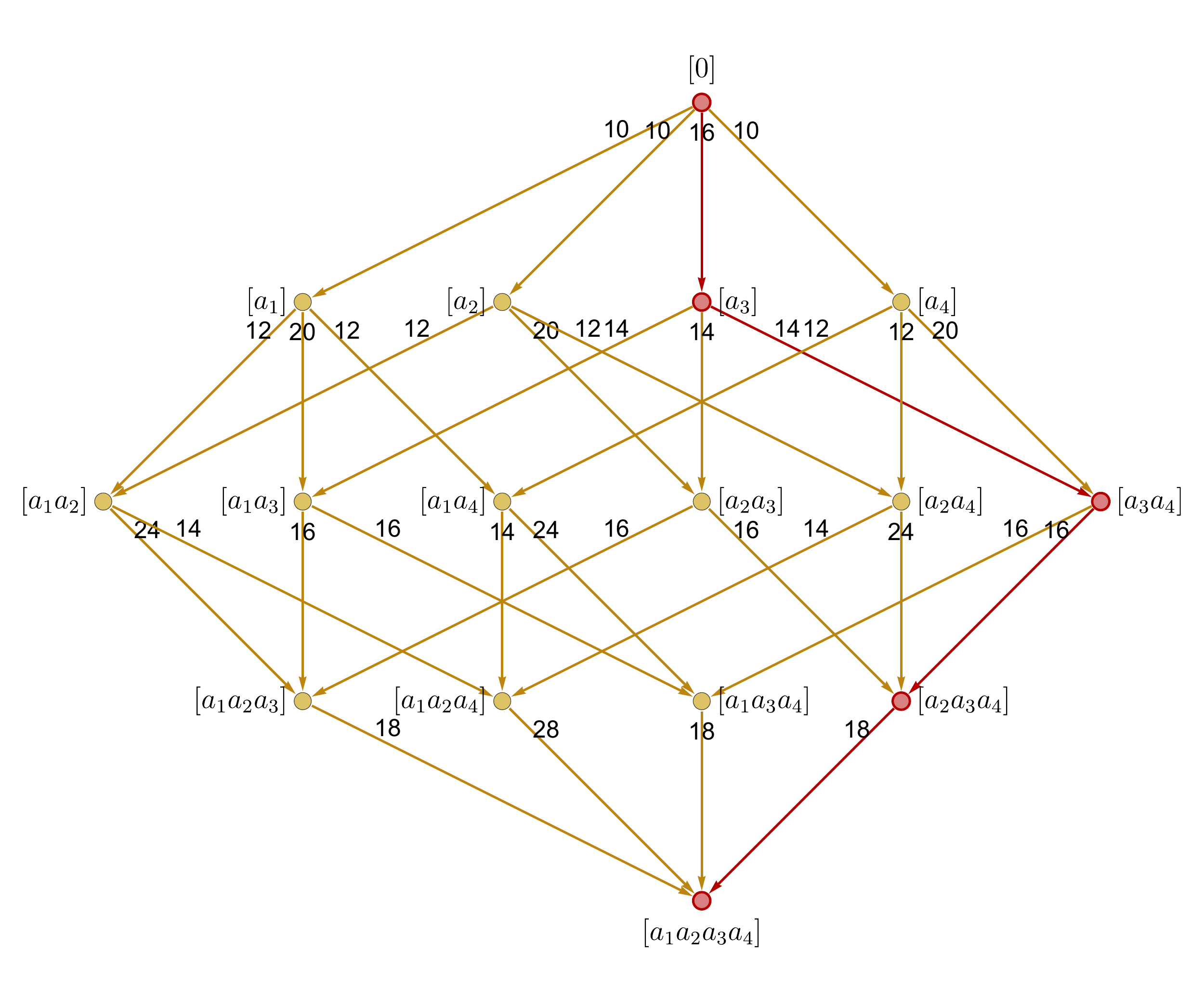}
\caption{
\label{fig:graph}
Graph representation of the VRRs for the 3-chain $f_{12}h_{23}$ (top left), cyclic $f_{12}g_{13}h_{23}$ (top right), 4-chain $f_{12}h_{23}i_{34}$ (bottom left) and trident $g_{13} h_{23}i_{34}$ (bottom right) operators. 
The edge label gives the number of terms in the corresponding VRR. 
The red path corresponds to the algorithm generating the smallest number of intermediates.}
\end{figure*}

VRR$_{4}^{\bA_2}$, VRR$_{4}^{\bA_3}$ and VRR$_{4}^{\bA_4}$ are used to build angular momentum over $\bA_{2}$, $\bA_{3}$ and $\bA_{4}$, respectively.
Their expressions are
\begin{equation}
\begin{split}
\label{eq:VRR4A2}
	\sexpval{\ba_{1} \ba_2^+ \ba_3 \ba_4}^{\bm{m}} 
	& = \bZ_2 \bA_2 \sexpval{\ba_1 \ba_2 \ba_3 \ba_4}^{\bm{m}} 
	+ \zeta_1 \zeta_3 \zeta_4 \bZ_{12} \sexpval{\ba_1 \ba_2 \ba_3 \ba_4}^{\{1\}}
	- \zeta_1 \zeta_3 \zeta_4 \bZ_{23} \sexpval{\ba_1 \ba_2 \ba_3 \ba_4}^{\{3\}} 
	\\
	& + \zeta_4 ( \zeta_1 \bZ_{12} - \zeta_3 \bZ_{23} ) \sexpval{\ba_1 \ba_2 \ba_3 \ba_4}^{\{5\}}
	+ \zeta_1 ( \zeta_3 + \zeta_4 ) \bZ_{12} \sexpval{\ba_1 \ba_2  \ba_3  \ba_4}^{\{6\}}
	\\
	& - \zeta_1 ( \zeta_3 \bZ_{23} + \zeta_4 \bZ_{24} ) \sexpval{\ba_1 \ba_2 \ba_3 \ba_4}^{\{8\}}
	+ ( \zeta_1 \bZ_{12} - \zeta_3 \bZ_{23} - \zeta_4 \bZ_{24} ) \sexpval{\ba_1 \ba_2 \ba_3 \ba_4}^{\{9\}}
	\\
	& + \frac{\ba_2}{2\zeta_2} \Big\{
	\sexpval{\ba_1 \ba_2^- \ba_3 \ba_4}^{\bm{m}} 
	- \zeta_1 \zeta_3 \zeta_4 \sexpval{\ba_1 \ba_2^- \ba_3 \ba_4}^{\{1\}}
	- \zeta_1 \zeta_3 \zeta_4 \sexpval{\ba_1 \ba_2^- \ba_3 \ba_4}^{\{3\}} 
	\\
	& - \zeta_4 ( \zeta_1 - \zeta_3 ) \sexpval{\ba_1 \ba_2^- \ba_3 \ba_4}^{\{5\}}
	+ \zeta_1 ( \zeta_3 + \zeta_4 ) \sexpval{\ba_1 \ba_2^- \ba_3  \ba_4}^{\{6\}}
	\\
	& - \zeta_1 ( \zeta_3 + \zeta_4 ) \sexpval{\ba_1 \ba_2^- \ba_3 \ba_4}^{\{8\}}
	- ( \zeta_1 + \zeta_3 + \zeta_4 ) \sexpval{\ba_1 \ba_2^- \ba_3 \ba_4}^{\{9\}}
	\Big\} 
	\\
	& + \frac{\ba_1}{2} \Big\{
	\zeta_3 \zeta_4 \sexpval{\ba_1^- \ba_2 \ba_3 \ba_4}^{\{1\}} 
	+ \zeta_4 \sexpval{\ba_1^- \ba_2 \ba_3 \ba_4}^{\{5\}}
	+ (\zeta_3 + \zeta_4) \sexpval{\ba_1^- \ba_2 \ba_3 \ba_4}^{\{6\}}
	+ \sexpval{\ba_1^- \ba_2 \ba_3 \ba_4}^{\{9\}}
	\Big\} 
	\\
	& + \frac{\ba_3}{2} \Big\{
	\zeta_1 \zeta_4 \sexpval{\ba_1 \ba_2 \ba_3^- \ba_4}^{\{3\}} 
	+ \zeta_4 \sexpval{\ba_1 \ba_2 \ba_3^- \ba_4}^{\{5\}}
	+ \zeta_1 \sexpval{\ba_1 \ba_2 \ba_3^- \ba_4}^{\{8\}}
	+ \sexpval{\ba_1 \ba_2 \ba_3^- \ba_4}^{\{9\}}
	\Big\} 
	\\
	& + \frac{\ba_4}{2} \Big\{
	\zeta_1 \sexpval{\ba_1 \ba_2 \ba_3 \ba_4^-}^{\{8\}}
	+ \sexpval{\ba_1 \ba_2 \ba_3 \ba_4^-}^{\{9\}}
	\Big\},
\end{split}
\end{equation}

\begin{equation}
\begin{split}
\label{eq:VRR4A3}
	\sexpval{\ba_{1} \ba_2 \ba_3^+ \ba_4}^{\bm{m}} 
	& = \bZ_3 \bA_3 \sexpval{\ba_1 \ba_2 \ba_3 \ba_4}^{\bm{m}} 
	+ \zeta_1 \zeta_2 \zeta_4 \bZ_{13} \sexpval{\ba_1 \ba_2 \ba_3 \ba_4}^{\{2\}}
	+ \zeta_1 \zeta_2 \zeta_4 \bZ_{23} \sexpval{\ba_1 \ba_2 \ba_3 \ba_4}^{\{3\}} 
	\\
	& + \zeta_1 \zeta_2 \zeta_4 \bZ_{34} \sexpval{\ba_1 \ba_2 \ba_3 \ba_4}^{\{4\}} 
	+ \zeta_4 ( \zeta_1 \bZ_{13} + \zeta_2 \bZ_{23} ) \sexpval{\ba_1 \ba_2  \ba_3  \ba_4}^{\{5\}}
	- \zeta_4 ( \zeta_1 + \zeta_2 ) \bZ_{34} \sexpval{\ba_1 \ba_2 \ba_3 \ba_4}^{\{6\}}
	\\
	& + \zeta_2 ( \zeta_1 \bZ_{13} - \zeta_4 \bZ_{34} ) \sexpval{\ba_1 \ba_2 \ba_3 \ba_4}^{\{7\}}
	+ \zeta_1 ( \zeta_2 \bZ_{23} - \zeta_4 \bZ_{34} ) \sexpval{\ba_1 \ba_2 \ba_3 \ba_4}^{\{8\}}
	\\
	& + ( \zeta_1 \bZ_{13} + \zeta_2 \bZ_{23} - \zeta_4 \bZ_{34} ) \sexpval{\ba_1 \ba_2 \ba_3 \ba_4}^{\{9\}}
	\\
	& + \frac{\ba_3}{2\zeta_3} \Big\{
	\sexpval{\ba_1 \ba_2 \ba_3^- \ba_4}^{\bm{m}} 
	- \zeta_1 \zeta_2 \zeta_4 \sexpval{\ba_1 \ba_2 \ba_3^- \ba_4}^{\{2\}}
	- \zeta_1 \zeta_2 \zeta_4 \sexpval{\ba_1 \ba_2 \ba_3^- \ba_4}^{\{3\}} 
	- \zeta_1 \zeta_2 \zeta_4 \sexpval{\ba_1 \ba_2 \ba_3^- \ba_4}^{\{4\}} 
	\\
	& - \zeta_4 ( \zeta_1 + \zeta_2 ) \sexpval{\ba_1 \ba_2  \ba_3^- \ba_4}^{\{5\}}
	- \zeta_4 ( \zeta_1 + \zeta_2 ) \sexpval{\ba_1 \ba_2 \ba_3^- \ba_4}^{\{6\}}
	 - \zeta_2 ( \zeta_1 + \zeta_4 ) \sexpval{\ba_1 \ba_2 \ba_3^- \ba_4}^{\{7\}}
	\\
	& - \zeta_1 ( \zeta_2  + \zeta_4 ) \sexpval{\ba_1 \ba_2 \ba_3^- \ba_4}^{\{8\}}
	- ( \zeta_1 + \zeta_2 + \zeta_4 ) \sexpval{\ba_1 \ba_2 \ba_3^- \ba_4}^{\{9\}}
	\Big\} 
	\\
	& + \frac{\ba_1}{2} \Big\{
	\zeta_2 \zeta_4 \sexpval{\ba_1^- \ba_2 \ba_3 \ba_4}^{\{2\}}
	+ \zeta_4 \sexpval{\ba_1^- \ba_2  \ba_3 \ba_4}^{\{5\}}
	+ \zeta_2 \sexpval{\ba_1^- \ba_2 \ba_3 \ba_4}^{\{7\}}
	+ \sexpval{\ba_1^- \ba_2 \ba_3 \ba_4}^{\{9\}}
	\Big\} 
	\\
	& + \frac{\ba_2}{2} \Big\{
	\zeta_1 \zeta_4 \sexpval{\ba_1 \ba_2^- \ba_3 \ba_4}^{\{3\}} 
	+ \zeta_4 \sexpval{\ba_1 \ba_2^-  \ba_3  \ba_4}^{\{5\}}
	+ \zeta_1 \sexpval{\ba_1 \ba_2^- \ba_3 \ba_4}^{\{8\}}
	+ \sexpval{\ba_1 \ba_2^- \ba_3 \ba_4}^{\{9\}}
	\Big\} 
	\\
	& + \frac{\ba_4}{2} \Big\{
	\zeta_1 \zeta_2 \sexpval{\ba_1 \ba_2 \ba_3 \ba_4^-}^{\{4\}} 
	+ ( \zeta_1 + \zeta_2 ) \sexpval{\ba_1 \ba_2 \ba_3 \ba_4^-}^{\{6\}}
	+ \zeta_2 \sexpval{\ba_1 \ba_2 \ba_3 \ba_4^-}^{\{7\}}
	+ \zeta_1 \sexpval{\ba_1 \ba_2 \ba_3 \ba_4^-}^{\{8\}}
	\\
	& + \sexpval{\ba_1 \ba_2 \ba_3 \ba_4^-}^{\{9\}}
	\Big\} 	
\end{split}
\end{equation}

\begin{equation}
\begin{split}
\label{eq:VRR4A4}
	\sexpval{\ba_{1} \ba_2 \ba_3 \ba_4^+}^{\bm{m}} 
	& = \bZ_4 \bA_4 \sexpval{\ba_1 \ba_2 \ba_3 \ba_4}^{\bm{m}} 
	+ \zeta_1 \zeta_2 \zeta_3 \bZ_{34} \sexpval{\ba_1 \ba_2 \ba_3 \ba_4}^{\{4\}}
	+ \zeta_3 ( \zeta_1 + \zeta_2 ) \bZ_{34} \sexpval{\ba_1 \ba_2  \ba_3  \ba_4}^{\{6\}}
	\\
	& + \zeta_2 ( \zeta_1 \bZ_{14} + \zeta_3 \bZ_{34} ) \sexpval{\ba_1 \ba_2 \ba_3 \ba_4}^{\{7\}}
	+ \zeta_1 ( \zeta_2 \bZ_{24} + \zeta_3 \bZ_{34} ) \sexpval{\ba_1 \ba_2 \ba_3 \ba_4}^{\{8\}}
	\\
	& + ( \zeta_1 \bZ_{14} + \zeta_2 \bZ_{24} + \zeta_3 \bZ_{34} ) \sexpval{\ba_1 \ba_2 \ba_3 \ba_4}^{\{9\}}
	\\
	& + \frac{\ba_4}{2\zeta_4} \Big\{
	\sexpval{\ba_1 \ba_2 \ba_3 \ba_4^-}^{\bm{m}} 
	- \zeta_1 \zeta_2 \zeta_3 \sexpval{\ba_1 \ba_2 \ba_3 \ba_4^-}^{\{4\}}
	- \zeta_3 ( \zeta_1 + \zeta_2 ) \sexpval{\ba_1 \ba_2  \ba_3  \ba_4^-}^{\{6\}}
	\\
	& - \zeta_2 ( \zeta_1 + \zeta_3 ) \sexpval{\ba_1 \ba_2 \ba_3 \ba_4^-}^{\{7\}}
	- \zeta_1 ( \zeta_2 + \zeta_3 ) \sexpval{\ba_1 \ba_2 \ba_3 \ba_4^-}^{\{8\}}
	- ( \zeta_1 + \zeta_2 + \zeta_3 ) \sexpval{\ba_1 \ba_2 \ba_3 \ba_4^-}^{\{9\}}
	\Big\} 	
	\\
	& + \frac{\ba_1}{2} \Big\{
	\zeta_2 \sexpval{\ba_1^- \ba_2 \ba_3 \ba_4}^{\{7\}}
	+ \sexpval{\ba_1^- \ba_2 \ba_3 \ba_4}^{\{9\}}
	\Big\} 	
	+ \frac{\ba_2}{2} \Big\{
	\zeta_1 \sexpval{\ba_1 \ba_2^- \ba_3 \ba_4}^{\{8\}}
	+ \sexpval{\ba_1 \ba_2^- \ba_3 \ba_4}^{\{9\}}
	\Big\} 	
	\\
	& + \frac{\ba_3}{2} \Big\{
	\zeta_1 \zeta_2 \sexpval{\ba_1 \ba_2 \ba_3^- \ba_4}^{\{4\}}
	+ ( \zeta_1 + \zeta_2 ) \sexpval{\ba_1 \ba_2  \ba_3^-  \ba_4}^{\{6\}}
	+ \zeta_2 \sexpval{\ba_1 \ba_2 \ba_3^- \ba_4}^{\{7\}}
	+ \zeta_1 \sexpval{\ba_1 \ba_2 \ba_3^- \ba_4}^{\{8\}}
	\\
	& + \sexpval{\ba_1 \ba_2 \ba_3^- \ba_4}^{\{9\}}
	\Big\} 	
\end{split}
\end{equation}
Again, the corresponding  expressions for VRR$_{1}$, VRR$_{2}$ , VRR$_{3}$ can be easily derived from Eqs.~\eqref{eq:VRR4A2}, \eqref{eq:VRR4A3} and \eqref{eq:VRR4A4}.
The number of terms for each of these VRRs is reported in Fig.~\ref{fig:graph} for the 3-chain $f_{12}h_{23}$ (top left), cyclic $f_{12}g_{13}h_{23}$ (top right), 4-chain $f_{12}h_{23}i_{34}$ (bottom left) and trident $g_{13} h_{23}i_{34}$ (bottom right) operators.

\subsection{
\label{sec:TRR}
Transfer recurrence relations}
TRRs redistribute angular momentum between centers hosting to different electrons.
Using translational invariance, one can derive 
\begin{equation}
\begin{split}
\label{eq:TRR}
	\sexpval{\ba_1^+ \ba_2 \ba_3  \ba_4 } 
	& 
	=\frac{\ba_1}{2\zeta_1} \sexpval{\ba_1^- \ba_2 \ba_3 \ba_4 }
	+\frac{\ba_2}{2\zeta_1} \sexpval{\ba_1 \ba_2^- \ba_3 \ba_4 }
	+\frac{\ba_3}{2\zeta_1} \sexpval{\ba_1 \ba_2 \ba_3^- \ba_4 }
	+\frac{\ba_4}{2\zeta_1} \sexpval{\ba_1 \ba_2 \ba_3 \ba_4^- }
	\\
	& 
	-\frac{\zeta_2}{\zeta_1} \sexpval{\ba_1 \ba_2^+ \ba_3 \ba_4 }
	-\frac{\zeta_3}{\zeta_1} \sexpval{\ba_1 \ba_2 \ba_3^+ \ba_4 }
	-\frac{\zeta_4}{\zeta_1} \sexpval{\ba_1 \ba_2 \ba_3 \ba_4^+ }
	\\
	& 
	-\frac{\beta_1\,\bA_1 \bB_1 + \beta_2\,\bA_2 \bB_2 + \beta_3\,\bA_3 \bB_3  + \beta_4\,\bA_4 \bB_4}{\zeta_1}
	\sexpval{\ba_1 \ba_2 \ba_3  \ba_4 }.
\end{split}
\end{equation}

\subsection{
\label{sec:HRR}
Horizontal recurrence relations}
The so-called HRRs enable to shift momentum between centers over the same electronic coordinate:
\begin{subequations}
\begin{gather}
	\braket{ \ba_1 \ba_2 \ba_3  \ba_4 }{  \bb_4^+ }
	= \braket{\ba_1 \ba_2 \ba_3  \ba_4^+  }{  \bb_4 }
	+ \bA_4 \bB_4 \braket{ \ba_1 \ba_2 \ba_3  \ba_4  }{  \bb_4 },
\label{eq:HRR_4}
	\\
	\braket{ \ba_1 \ba_2 \ba_3  \ba_4  }{  \bb_3^+  \bb_4 }
	= \braket{ \ba_1 \ba_2 \ba_3^+  \ba_4  }{ \bb_3  \bb_4 } 
	+ \bA_3 \bB_3 \braket{ \ba_1 \ba_2 \ba_3  \ba_4  }{  \bb_3  \bb_4 },
\label{eq:HRR_3}
	\\
	\braket{ \ba_1 \ba_2 \ba_3  \ba_4  }{  \bb_2^+ \bb_3  \bb_4 }
	= \braket{ \ba_1 \ba_2^+ \ba_3  \ba_4 }{  \bb_2 \bb_3  \bb_4 }
	+ \bA_2 \bB_2 \braket{ \ba_1 \ba_2 \ba_3  \ba_4  }{  \bb_2 \bb_3  \bb_4 },
\label{eq:HRR_2}
	\\
	\braket{ \ba_1 \ba_2 \ba_3  \ba_4 }{  \bb_1^+ \bb_2 \bb_3  \bb_4 }
	= \braket{ \ba_1^+ \ba_2 \ba_3  \ba_4  }{ \bb_1 \bb_2 \bb_3  \bb_4 }
	+ \bA_1 \bB_1 \braket{ \ba_1 \ba_2 \ba_3  \ba_4 }{  \bb_1 \bb_2 \bb_3  \bb_4 }.
\label{eq:HRR_1}
\end{gather}
\end{subequations}
Note that HRRs can be applied to contracted integrals because they are independent of the contraction coefficients and exponents.

\begin{figure}
	\includegraphics[width=0.49\linewidth]{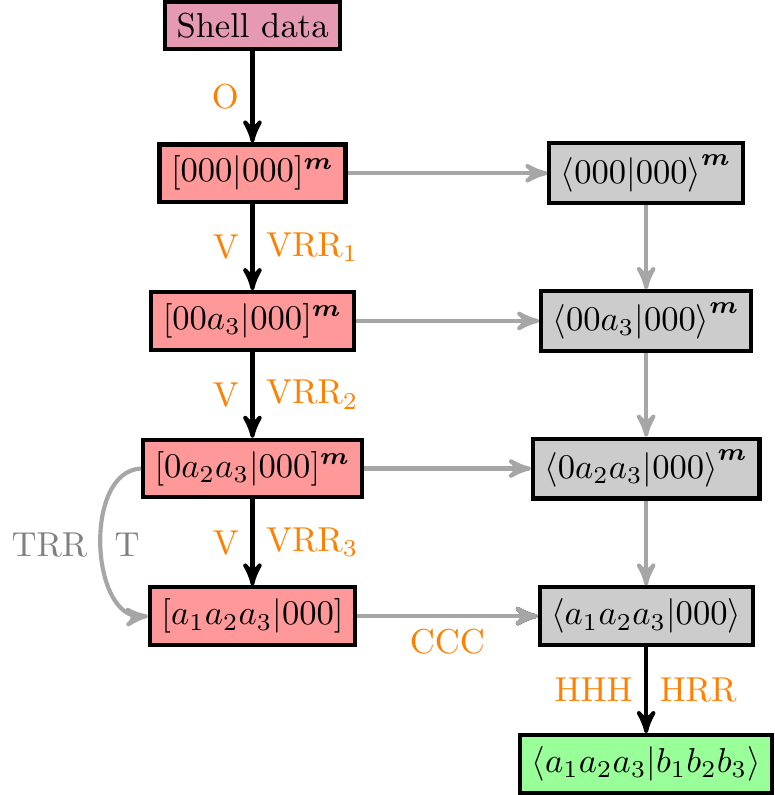}
	\includegraphics[width=0.49\linewidth]{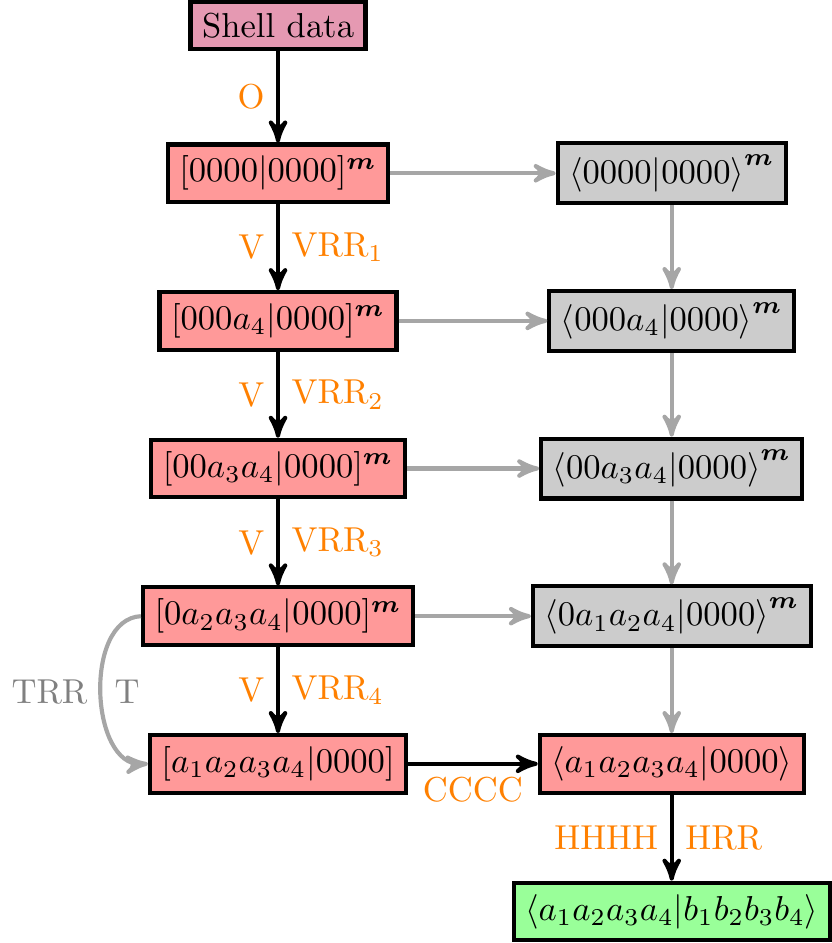}
\caption{
\label{fig:algo}
Schematic representation of the algorithm used to compute three-electron integrals over the 3-chain operator $f_{12}g_{23}$ (left) and the four-electron integrals over the 4-chain operator $f_{12}h_{23}i_{34}$ (right).
The three- and four-electron algorithms follow a OVVVCCCHHH and OVVVVCCCCHHHH path, respectively.}
\end{figure}

\section{
\label{sec:algo}
Algorithm}
In this Section, we describe a recursive scheme for the computation of three- and four-electron integrals based on a late contraction scheme \`a la Head-Gordon-Pople (HGP). \cite{HGP}
The general skeleton of the algorithm is shown in Fig.~\ref{fig:algo} for two representative examples: the 3-chain operator $f_{12}g_{23}$ (left) and the 4-chain operator $f_{12}h_{23}i_{34}$ (right).
First, let us focus on the 3-chain operator.

To compute a class of three-electron integrals $\braket{a_1 a_2 a_3}{b_1 b_2 b_3}$, starting from the fundamental integrals $\sexpval{0 0 0}^{\bm{m}}$, we first build up angular momentum over center $\bA_3$ with the 6-term VRR$_1$ to obtain $\sexpval{0 0 a_3}$. 
Then, we use the 10-term VRR$_2$ over $\bA_2$ to obtain $\sexpval{0 a_2 a_3}$. 
Finally, we build up momentum over the last bra center $\bA_1$ using the 9-term VRR$_3$ to get $\sexpval{a_1 a_2 a_3}$.

Note that, in our previous paper, \cite{3ERI1} we claimed that it would be computationally cheaper to use the 6-term TRR instead of VRR$_3$ because the number of terms in the TRR is much smaller than in VRR$_3$.
However, we have found here that the number of intermediates (i.e.~the number of pre-computed classes needed to calculate a given class) required by the paths involving the TRR is much larger (see Table \ref{tab:RRcount}).
This can be easily understood noticing that, to increase the momentum by one unit on the last center, one must increase the momentum by the same amount on all the other centers (as evidenced by the second term in the right-hand side of \eqref{eq:TRR}). 
Hence, the TRR is computationally expensive for three- and four-electron integrals due to the large number of centers.

As illustrated in the top left graph of Fig.~\ref{fig:graph}, other paths, corresponding to different VRRs, are possible.
However, we have found that they do generate a larger number of intermediates, as reported in Table \ref{tab:RRcount}.

Similarly to the 3-chain operator, for the 4-chain operator, we get $\sexpval{a_1 a_2 a_3 a_4}$ by successively building up momentum over $\bA_4$, $\bA_3$, $\bA_2$ and $\bA_1$.
The number of intermediates required by the other paths are gathered in Table \ref{tab:RRcount}.
Again, the paths involving the 8-term TRR (reported in Eq.~\eqref{eq:TRR}) are much more expensive.

The last two steps of the algorithm are common to the three- and four-electron integral schemes.
Following the HGP algorithm, \cite{HGP} we contract the integrals: $\sbraket{\ba_1 \ba_2 \ba_3  \ba_4}{\bO \bO \bO \bO} $ to form $\braket{\ba_1 \ba_2 \ba_3  \ba_4}{\bO \bO \bO \bO}$ in the four-electron case, or $\sbraket{\ba_1 \ba_2 \ba_3 }{\bO \bO \bO} $ to form $\braket{\ba_1 \ba_2 \ba_3}{\bO \bO \bO}$ in the three-electron case.
More details about the contraction step can be found in Ref.~\onlinecite{Gill94b}.
The final step of the algorithm shifts momentum to the ket centers from the bra centers with the help of the 2-term HRRs reported in Sec.~\ref{sec:HRR}.

\begin{table*}
\caption{
\label{tab:RRcount}
Number of intermediates required to compute various integral classes for two-, three- and four-electron operators.
The path generating the minimum number of intermediates is highlighted in bold.
The number of terms in the RRs and the associated incremental center are also reported.
} 
\begin{ruledtabular}
\begin{tabular}{llllllrrr}
Integral			&	 type		&	operator				&	path		& number				& centers								&	\mc{3}{c}{integral class}		\\
																						\cline{7-9}
				&	 		&						&			& of terms				&									&	$\sexpval{p \ldots p}$	&	$\sexpval{d \ldots d}$	&	$\sexpval{f \ldots f}$		\\
\hline
two-electron		&	   		&	$f_{12}$				&	\bf VV	&	\bf (4,6)			&	($\bA_2$,$\bA_1$)					&	\bf 4		&	\bf 13		&	\bf 25		\\	
				&	   		&						&	VT		&	(4,4)				&	($\bA_2$,$\bA_1$)					&	7		&	19			&	37			\\
\hline
three-electron		&	chain   	&	$f_{12}h_{23}$			&	VVV		&	(6,7,12)			&	($\bA_3$,$\bA_1$,$\bA_2$)			&	34		&	230			&	881			\\					
				&		  	&						&	\bf VVV	&	\bf (6,10,9)		&	($\bA_3$,$\bA_2$,$\bA_1$)			&	\bf 32	&	\bf 209		&	\bf 778		\\
				&		  	&						&	VVV		&	(8,8,9)			&	($\bA_2$,$\bA_3$,$\bA_1$)			&	32		&	212			&	801			\\					
				&		  	&						&	VVT		&	(6,7,6)			&	($\bA_3$,$\bA_1$,$\bA_2$)			&	38		&	246			&	873			\\					
				&		  	&						&	VVT		&	(6,10,6)			&	($\bA_3$,$\bA_2$,$\bA_1$)			&	43		&	314			&	1,256			\\					
				&		  	&						&	VVT		&	(8,8,6)			&	($\bA_2$,$\bA_3$,$\bA_1$)			&	40		&	260			&	923			\\					
\hline
				&	cyclic	&	$f_{12}g_{13}h_{23}$	&	\bf VVV	&	\bf (8,10,12)		&	($\bA_3$,$\bA_2$,$\bA_1$)			&	\bf 52	&	\bf 469	 	&	\bf 2,216		\\	
				&		   	&						&	VVT		&	(8,10,6)			&	($\bA_3$,$\bA_2$,$\bA_1$)			&	61		&	539			&	2,426		\\					
\hline
four-electron		&	chain   	&	$f_{12}h_{23}i_{34}$		&	VVVV	&	(10,11,20,24)		&	($\bA_4$,$\bA_1$,$\bA_3$,$\bA_2$)	&	465		&	13,781		&	150,961		\\					
				&		   	&						&	\bf VVVV	&	\bf (10,18,20,17)	&	($\bA_4$,$\bA_3$,$\bA_2$,$\bA_1$)	&	\bf 436	&	\bf 12,535		&	\bf 133,891	\\					
				&		   	&						&	VVVV	&	(14,14,20,17)&	($\bA_3$,$\bA_4$,$\bA_2$,$\bA_1$)	&	433		&	12,704		&	138,913		\\
				&		   	&						&	VVVV	&	(14,18,16,17)		&	($\bA_3$,$\bA_2$,$\bA_4$,$\bA_1$)	&	435		&	12,863		&	141,679		\\
				&		   	&						&	VVVT	&	(10,11,20,8)		&	($\bA_4$,$\bA_1$,$\bA_3$,$\bA_2$)	&	532		&	16,295		&	181,178		\\								
				&		   	&						&	VVVT	&	(10,18,20,8)		&	($\bA_4$,$\bA_3$,$\bA_2$,$\bA_1$)	&	560		&	17,029		&	188,242			\\					
				&		   	&						&	VVVT	&	(14,14,20,8)	&	($\bA_3$,$\bA_4$,$\bA_2$,$\bA_1$)		&	559		&	17,487		&	199,050			\\
				&		   	&						&	VVVT	&	(14,18,16,8)		&	($\bA_3$,$\bA_2$,$\bA_4$,$\bA_1$)	&	543		&	16,612		&	185,869			\\
\hline
				&	trident	 &	$g_{13} h_{23}i_{34}$	&	VVVV	&	(10,12,14,28)		&	($\bA_4$,$\bA_2$,$\bA_1$,$\bA_3$)	&	445		&	13,139		&	143,619		\\					
				&		   	&						&	VVVV	&	(10,12,24,18)		&	($\bA_4$,$\bA_2$,$\bA_3$,$\bA_1$)	&	447		&	13,381		&	148,911		\\					
				&	   		&						&	VVVV	&	(10,20,16,18)		&	($\bA_4$,$\bA_3$,$\bA_2$,$\bA_1$)	&	418		&	12,447		&	133,853		\\	
				&		   	&						&	\bf VVVV	&	\bf (16,14,16,18)	&	($\bA_3$,$\bA_4$,$\bA_2$,$\bA_1$)	&	\bf 418	&	\bf 12,054		&	\bf 129,322	\\
				&		   	&						&	VVVT	&	(10,12,14,8)		&	($\bA_4$,$\bA_2$,$\bA_1$,$\bA_3$)	&	470		&	13,306		&	136,584		\\					
				&		   	&						&	VVVT	&	(10,12,24,8)		&	($\bA_4$,$\bA_2$,$\bA_3$,$\bA_1$)	&	546			&16,917		& 	191,171	\\			
				&	   		&						&	VVVT	&	(10,20,16,8)		&	($\bA_4$,$\bA_3$,$\bA_2$,$\bA_1$)	&	521			&15,515		&	168,958	\\	
				&		   	&						&	VVVT	&	(16,14,16,8)		&	($\bA_3$,$\bA_4$,$\bA_2$,$\bA_1$)	&	499			& 13,969		&	142,264	\\
\end{tabular}
\end{ruledtabular}	
\end{table*}

\section{Concluding remarks}
In this study, we have reported recurrence relations (RRs) for the efficient and accurate computation of four-electron integrals over Gaussian basis functions and a general class of multiplicative four-electron operators of the form $f_{12} g_{13}h_{23}i_{34}$. Starting from this master operator, one can easily derive the RRs for various operators arising in explicitly-correlated methods following simple diagrammatic rules (see Fig.~\ref{fig:tree}).

Here, we have derived three types of RRs: i) starting from the fundamental integrals, vertical RRs (VRRs) allow to increase the angular momentum over the bra centers; ii)
the transfer RR (TRR) redistributes angular momentum between centers hosting different electrons, and can be used instead of the VRR on the last bra center; iii)
the horizontal RRs (HRRs) enable to shift momentum from the bra to the ket centers corresponding to the same electronic coordinate. 
Importantly, HRRs can be applied to contracted integrals.

Finally, after carefully studying the different paths one can follow to build up angular momentum (see Fig.~\ref{fig:graph}), we have proposed a late-contraction recursive scheme which minimizes the number of intermediates to be computed  (see Fig.~\ref{fig:algo}).
We believe our approach represents a major step towards a fast and accurate computational scheme for three- and four-electron integrals within explicitly-correlated methods. \alert{It also paves the way to contraction-effective methods for these types of integrals. \cite{GG16} In particular, an early contraction scheme would have significant computational benefits.}

\begin{acknowledgments}
P.F.L.~thanks the NCI National Facility for generous grants of supercomputer time, and the Australian Research Council for a Discovery Project grant (DP140104071).
\end{acknowledgments}

%

\end{document}